\documentstyle[preprint,aps]{revtex}
\begin{document}
\draft
\preprint{\vbox{
\hbox{IFT-P.003/94}
\hbox{hep-ph/9401272}
\hbox{January 1994}
}}
%\documentstyle[preprint,revtex,eqsecnum]{aps}
%\flushbottom
\title{
$SU(4)_L \otimes U(1)_N$ model for the electroweak interactions}
\author{ F. Pisano and V. Pleitez}
\address{
Instituto de F\'\i sica Te\'orica\\
Universidade Estadual Paulista\\
Rua Pamplona, 145\\
01405-900-- S\~ao Paulo, SP\\
Brazil}
\maketitle
\begin{abstract}
Assuming the existence of right-handed neutrinos, we consider an
electroweak model based on the gauge symmetry  $SU(4)_L\otimes
U(1)_N$. We study the
neutral currents coupled to all neutral vector bosons present in the
theory. There are no flavor changing neutral currents at tree level,
coupled with the lightest neutral vector boson.

\end{abstract}
\narrowtext
Symmetry principles have been used in elementary particle physics at
least since the discovery of the neutron. A symmetry is useful to
both issues: the classification of particles and the dynamics of the
interactions among them. The point is that there must be a part of
the particle spectrum in which the symmetry manifests itself at
least in an approximate way. This is the case for quarks $u$ and $d$
and in the leptonic sector for the electron-neutrino  and
electron. For instance, the $SU(2)$ appears as an
approximate symmetry in the doublets $(\nu_e,e)^T$. If one assumes
this symmetry among these particles and in the sequential families as
well, almost all the model's predictions are determined.

The full symmetry of the so called Standard Model is the gauge group
$SU(3)_c\otimes SU(2)_L\otimes U(1)_Y$. This model spectacularly
explains all the available experimental data~\cite{pdg}. Usually it is
considered that this symmetry emerges at low energies as a result of
the breaking of higher symmetries. Probably, these huge symmetries are
an effect of grand unified scenarios and/or their
supersymmetric extensions.

Considering the lightest particles of the model as the sector in which
a symmetry is manifested, it is interesting that the lepton sector could
be the part of the model determining new approximate symmetries. For
instance, $\nu,e$ and $e^c$ could be in the same triplet of an
$SU(3)_L\otimes U(1)_N$ symmetry. This sort of model has been
proposed recently~\cite{pp}. In this case neutrinos can remain
massless in arbitrary order in perturbation theory, or they get a mass in
some modifications of the
models~\cite{ppt}. If we admit that right-handed neutrinos do exist,
it is possible to build a model in which $\nu^c,\nu$ and $e$ are in the same
multiplet of $SU(3)$~\cite{mpp}. In fact, if right-handed neutrinos are
introduced it is a more interesting
possibility to have $\nu,e,\nu^c$ and $e^c$ in the same multiplet
of a $SU(4)_L\otimes U(1)_N$ electroweak theory.

Notice that using the lightest leptons as the particles which
determine the approximate symmetry, {\em if each
generation is treated separately}, $SU(4)$ is the highest symmetry
group to be considered in the electroweak sector. A model with the
$SU(4)\otimes U(1)$ symmetry in the lepton sector was suggested some
years ago in Ref.~\cite{voloshin}. However, quarks were not
considered there. This symmetry in both, quarks and leptons, was pointed
out recently~\cite{vp} and here we will consider the details of such a
model.

Hence, our model has the full symmetry $SU(3)_c\otimes
SU(4)_L\otimes U(1)_N$.  These sort of models are
anomaly free only if there are equal number of ${\bf4}$ and
${\bf4}^*$ (considering the color degrees of freedom), and
furthermore requiring
the sum of all fermion charges to vanish. Two of the three quark
generations transform identically and one generation, it does not
matter which one, transforms in a different representation of
$SU(4)_L  \otimes U(1)_N$\cite{fn1}. This means that in these models
as in the
$SU(3)_c\otimes SU(3)_L\otimes U(1)_N$ ones~\cite{pp},  in order
to cancel anomalies, the number of families $(N_f)$ must be divisible
by the number of color degrees of freedom ($n$). Hence the simplest
alternative is $n=N_f=3$. On the other hand, at low energies these models
are indistinguishable from the Standard Model.

The electric charge operator is defined as
\begin{equation}
Q=\frac{1}{2}(\lambda_3-\frac{1}{\sqrt3}\lambda_8-\frac{2}{3}{\sqrt6}
\lambda_{15})+N,
\label{q}
\end{equation}
where the $\lambda$-matrices are a slightly modified version of the
usual ones~\cite{da},
\[\lambda_3=diag(1,-1,0,0),\;\lambda_8=(\frac{1}{\sqrt{3}})diag(1,1,-2,0),
\;\lambda_{15}=(\frac{1}{\sqrt{6}})diag(1,1,1,-3).\]

Leptons transform as $({\bf1},{\bf4},0)$, one generation, say
$Q_{1L}$, transforms as
$({\bf3},{\bf4},+2/3)$ and the other
two quark families, say $Q_{\alpha L},\,\alpha=2,3$, transform as
$({\bf3},{\bf4}^*,-1/3)$,
\begin{equation}
f_{aL}=\left(\begin{array}{c}
\nu_a\\
l_a\\
\nu^c_a\\
l^c_a
\end{array}\right)_L, \qquad Q_{1L} = \left(
\begin{array}{c}
u_1\\
d_1\\
u'\\
J
\end{array}\right)_L, \qquad Q_{\alpha L} = \left(
\begin{array}{c}
j_i\\
d'_i\\
u_\alpha\\
d_\alpha
\end{array}\right)_L,
\label{lq}
\end{equation}
where $a=e,\mu,\tau$; $u'$ and $J$ are new quarks with charge $+2/3$
and $+5/3$ respectively;
$j_i$ and $d'_i$, $i=1,2$ are new quarks with charge $-4/3$ and
$-1/3$ respectively. We remind that in Eq.~(\ref{lq})
all fields are still symmetry eigenstates.
Right-handed quarks transform as singlets under $SU(4)_L\otimes U(1)_N$.

Quark masses are generated by introducing the following Higgs
$SU(3)_c\otimes SU(4)_L\otimes U(1)_N$ multiplets: $\eta\sim({\bf1},{\bf4},0)$,
$\rho\sim({\bf1},{\bf4},+1/3)$ and $\chi\sim({\bf1},{\bf4},+1)$.
\begin{equation}
\eta=\left(
\begin{array}{c}
\eta_1^0 \\ \eta^-_1 \\ \eta^0_2 \\ \eta^+_2
\end{array}
\right)
\rho=\left(
\begin{array}{c}
\rho_1^+ \\ \rho^0 \\ \rho^+_2 \\ \rho^{++}
\end{array}
\right)
\chi=\left(
\begin{array}{c}
\chi_1^- \\ \chi^{--} \\ \chi^-_2 \\ \chi^0
\end{array}
\right).
\label{higgs4}
\end{equation}
In order to obtain massive charged leptons it is necessary to
introduce a $({\bf1},{\bf10}^*,0)$ Higgs multiplet, because the lepton mass
term transforms as
$\bar f^c_Lf_L\sim ({\bf6}_A\oplus {\bf10}_S)$.
The ${\bf6}_A$ will leave some leptons massless and some others
degenerate. Therefore we will choose the $H={\bf10}_S$. Explicitly
\begin{equation}
H=\left(
\begin{array}{cccc}
H^0_1\, &\, H^+_1\, &\, H_2^0\, &\, H_2^- \\
H^+_1\, &\, H_1^{++}\, &\, H_3^+\, &\, H_3^0 \\
H^0_2\, &\, H^+_3\, &\, H^0_4\, &\, H^-_4 \\
H^-_2\, &\, H^0_3\, &\, H^-_4\, &\, H^{--}_2
\end{array}
\right).
\label{10}
\end{equation}
If $\langle H^0_3\rangle \not=0$, $\langle H^0_{1,2,4}\rangle=0$ the
charged leptons get a mass but
neutrinos remain massless, at least at tree level.
In order to avoid mixing among primed and unprimed
quarks we can introduce another multiplet $\eta'$ transforming as
$\eta$ but with different vacuum expectation value (VEV). The
corresponding VEVs are the following
$\langle\eta\rangle=(v,0,0,0)$, $\langle\rho\rangle=(0,u,0,0)$,
$\langle\eta'\rangle=(0,0,v',0)$,
$\langle\chi\rangle=(0,0,0,w)$,
and $\langle H\rangle_{24}=\langle H_3^0\rangle=v''$ for the
decuplet. In this way we have
that the symmetry breaking of the $SU(4)_L\otimes U(1)_N$ group down
to $SU(3)_L\otimes
U(1)_{N'}$ is induced by the $\chi$ Higgs. The $SU(3)_L\otimes U(1)_{N'}$
symmetry is broken down into $U(1)_{em}$ by the $\rho,\eta$, $\eta'$
and $H$ Higgs.

The Yukawa interactions are
\begin{eqnarray}
-{\cal L}_Y&=&\frac{1}{2}G_{ab}\overline{f_{aL}^c}f_{bL}H+
F_{1k}\bar Q_{1L}u_{kR}\eta+
F_{\alpha k}\bar Q_{\alpha L}u_{kR}\rho^*\nonumber \\ & &\mbox{}
+F'_{1k}\bar Q_{1L}d_{kR}\rho+
F'_{\alpha k}\bar Q_{\alpha L}d_{kR}\eta^*+
h_1\bar Q_{1L}u'_R\eta'+h_{\alpha i}\bar Q_{\alpha L}d'_{i
R}\eta'^* \nonumber \\
& &\mbox{}
+\Gamma_1\bar Q_{1L}J_R\chi+\Gamma_{\alpha i}\bar Q_{\alpha
L}j_{i L}\chi^*+H.c.,
\label{yukawa}
\end{eqnarray}
where $a=e,\mu,\tau$; $k=1,2,3$; $i=1,2$ and $\alpha,\beta=2,3$.
We recall that up to now all fields are weak
eigenstates.

The electroweak gauge bosons of this theory consist of a
${\bf15}$ $W^i_\mu$, $i=1,...,15$ associated with $SU(4)_L$ and
a singlet $B_\mu$ associated with $U(1)_N$.

The gauge bosons $-\sqrt{2}W^+=W^1-iW^2$,
$-\sqrt{2}V^-_1=W^6-iW^7$, $-\sqrt{2}V_2^-=W^9-iW^{10}$,
$-\sqrt{2}V_3^-=W^{13}-iW^{14}$,
$-\sqrt{2}U^{--}=W^{11}-iW^{12}$ and $\sqrt{2}X^0=W^4+iW^5$ have masses

\begin{mathletters}
\label{wmass}
\begin{equation}
M^2_W=\frac{g^2}{4}(v^2+u^2+2v''^2), \quad
M^2_{V_1}=\frac{g^2}{4}(v'^2+u^2+2v''^2),
\label{a1}
\end{equation}
\begin{equation}
M^2_{V_2}=\frac{g^2}{4}(v^2+w^2+2v''^2),\quad
M^2_{V_3}=\frac{g^2}{4}(v'^2+w^2+2v''^2),
\label{b2}
\end{equation}
\begin{equation}
M^2_X= \frac{g^2}{8}(v^2+v'^2),  \quad
M^2_U= \frac{g^2}{4}(v^2+w^2+4v''^2).
\label{c1}
\end{equation}
\end{mathletters}
The mass matrix for the neutral vector bosons (up to a factor
$g^2/4$) in the $W^3,W^8,W^{15},B$ basis is
\begin{equation}
\left(
\begin{array}{llll}
v^2\!+\!u^2\!+\!v''^2\, &\, \frac{1}{\sqrt3}(v^2\!-\!u^2\!-\!v''^2)\, &\,
\frac{1}{\sqrt6}(v^2\!-\!u^2\!+\!2v''^2)\, &\, -2tu^2 \\
 \frac{1}{\sqrt3}(v^2\!-\!u^2\!-\!v''^2)\, &\,
\frac{1}{3}(v^2\!+\!4v'^2\!+\!u^2\!+\!v''^2)\, &
\frac{1}{3\sqrt2}(v^2\!-\!2v'^2\!+\!u^2\!-\!2v''^2)\, &\,
\frac{2}{\sqrt3}tu^2 \\
\frac{1}{\sqrt6}(v^2\!-\!u^2\!+\!2v''^2)\, &\,
\frac{1}{3\sqrt2}(v^2\!-\!2v'^2\!+\!u^2\!-\!2v''^2)
&\, \frac{1}{6}(v^2+v'^2+u^2+9w^2+4v''^2)\, &\,
\frac{2}{\sqrt6}t(u^2+3w^2) \\
-2tu^2\, &\, \frac{2}{\sqrt3}tu^2\, &\, \frac{2}{\sqrt6}t(u^2\!+\!3w^2)\,
&4t^2(u^2\!+\!w^2)
\end{array}
\right)
\label{zmass}
\end{equation}
where $t\equiv g'/g$. The matrix in (\ref{zmass}) has determinant
equal to zero as it must be
in order to have a massless photon. There are four neutral bosons: a
massless $\gamma$ and three massive
ones: $Z,Z',Z''$ such that $M_Z<M_{Z'}<M_{Z''}$. The lightest one,
say $Z$, corresponds to the neutral boson of the Standard Model.

The photon field is
\begin{equation}
A_\mu=\frac{1}{(1+4t^2)^{\frac{1}{2}}}\left(tW^3_\mu-\frac{t}{\sqrt3}W^8_\mu-
\frac{2\sqrt6}{3}tW^{15}_\mu+B_\mu
\right),
\label{foton}
\end{equation}
with the electric charge defined as
\begin{equation}
\vert
e\vert=\frac{gt}{(1+4t^2)^{\frac{1}{2}}}=\frac{g'}{(1+4t^2)^{\frac{1}{2}}}.
\label{charge}
\end{equation}
In the following, we will use the approximation
$v=u=v''\equiv v_1\ll v'=w\equiv v_2$.
In this approximation the three nonzero masses are given by\cite{mdt}
\begin{equation}
M^2_n\approx\frac{g^2}{4}(4\lambda_{n})v_2^2,\quad n=0,1,2;
\label{masszs}
\end{equation}
where
\begin{mathletters}
\label{lambda}
\begin{equation}
\lambda_{n}=\frac{1}{3}\left[A+2\left(A^2+3B\right)^{\frac{1}{2}}
\cos\left(\frac{2n\pi+\Theta}{3} \right)
 \right],
\label{ln}
\end{equation}
\begin{equation}
A=\frac{3}{4}+t^2+\left(\frac{5}{4}+t^2\right)a^2,\quad
B=-\frac{1}{8}(1+3t^2)-\frac{1}{4}(3+7t^2)a^2,
\label{ab}
\end{equation}
\begin{equation}
C=\frac{3}{32}(1+4t^2)a^2,\quad \Theta=\arccos\left[
\frac{2A^3+9AB+27C}{2(A^2+3B)^{\frac{3}{2}}}\right],
\label{c}
\end{equation}
\end{mathletters}
and we have defined $a\equiv v_1/v_2$. The respective eigenvectors
are
\begin{equation}
Z_{n\mu}\approx x_nW^3_\mu+y_nW^8_\mu+z_nW^{15}_\mu+w_nB,
\label{autovec}
\end{equation}
with
\begin{mathletters}
\label{xyzw}
\begin{equation}
x_n=-\frac{a^2}{2t}\cdot\frac{4\lambda_n-[2(a^2+3)t^2+a^2+1]}
{(a^2-1)(2\lambda_n-a^2)}\,w_n,
\label{1a}
\end{equation}
\begin{eqnarray}
y_n&=&\frac{\sqrt3}{6t(a^2-1)(2\lambda_n-a^2)}
\left[32\lambda_n^2-4[8a^2+3+8(a^2+1)t^2]\lambda_n\right.
\nonumber \\ & &\mbox{}
\left.+[5a^2+9+2(11a^2+17)t^2]a^2\right]w_n,
\label{2a}
\end{eqnarray}
\begin{equation}
z_n=-\frac{2}{\sqrt6t(a^2-1)}\cdot [2\lambda_n-2(a^2+1)t^2-a^2]\,w_n,
\label{3a}
\end{equation}
\begin{equation}
w^2_n=\frac{1}{1+x_n^2/w_n^2+y_n^2/w_n^2+z_n^2/w_n^2}.
\label{4a}
\end{equation}
\end{mathletters}
The hierarchy of the masses is
$M_0>M_2>M_1$. Hence we can identify the eigenvector with
$n=1$ as being the neutral vector boson of the Standard Model.

However, we have checked numerically if the respective eigenvalue
satisfies the relation $M^2_Z/M^2_W=1/\cos^2_W$, with
$\theta_W$
the weak mixing angle. Using  $a=0.01$ and
$t=1.79$\cite{fn} we
obtain $M_Z^2/M^2_W\approx0.97$ which does not agree with the value
of the Standard Model $M_Z^2/M^2_W\approx1.30$ with
$\sin^2\theta_W=0.2325$. In fact,
$M_Z/M_W\approx 0.99$. This suggests that, in this model, $Z$ and $W$
are mass degenerate at tree level. Hence, the right value for the
ratio $M_Z/M_W$ must arise only through radiative corrections.
We recall that in models with
$SU(3)_L\otimes U(1)_N$
symmetry $M_Z/M_W$ is bounded from above and
the weak mixing angle has an upper bound.

For these values of $a$ and $t$ the other two neutral bosons have
$M_{Z'}/M_W\approx 60$ and $M_{Z''}/M_W\approx 189$.
If $a<0.1$, all these results depend very weakly
on the value chosen for $a$.

The weak neutral currents have been, up to now, an important test of
the Standard Model. In particular it has been possible to determine
the fermion couplings, so far all experimental data are in agreement
with the model. In the present model, the
neutral currents couple to the $Z_n$ neutral boson as follows
\begin{equation}
{\cal L}_n^{NC}=-\frac{g}{2c_W}[\bar\psi_L\gamma^\mu\psi_LL_\psi
+\bar\psi_R\gamma^\mu\psi_RR_\psi]Z_{n\mu}
\label{nc}
\end{equation}
where $c_W\equiv \cos\theta_W$ and
\begin{mathletters}
\label{gim1}
\begin{equation}
L^n_{u_1}=-c_W\left(x_n+\frac{1}{\sqrt3}y_n
+\frac{1}{\sqrt6}z_n+\frac{4}{3}w_nt \right),
\label{gim1a}
\end{equation}
\begin{equation}
L^n_{u_\alpha}=-c_W\left(
x_n-\frac{1}{\sqrt3}y_n-\frac{1}{\sqrt6}z_n-\frac{2}{3}w_nt\right),
\label{gim1b}
\end{equation}
\begin{equation}
L^n_{u'}=-c_W\left(-\frac{2}{\sqrt3}y_n
+\frac{1}{\sqrt6}z_n+\frac{4}{3}w_nt\right),
\label{gim1c}
\end{equation}
\end{mathletters}
\begin{equation}
R^n_{u_1}=R^n_{u_\alpha}=R^n_{u'}=
-\frac{4}{3}\,c_W\,w_n\,t
\label{gim2}
\end{equation}
for the charge 3/2 quarks, and
\begin{mathletters}
\label{gim3}
\begin{equation}
L^n_{d_1}=-c_W\left(-x_n+\frac{1}{\sqrt3}
y_n+\frac{1}{\sqrt6}z_n+\frac{4}{3}w_nt\right),
\label{gim3a}
\end{equation}
\begin{equation}
L^n_{d_\alpha}=-c_W\left(-x_n
-\frac{1}{\sqrt3}y_n-\frac{1}{\sqrt6}z_n-\frac{2}{3}w_nt\right),
\label{gim3b}
\end{equation}
\begin{equation}
L^n_{d'_i}=-c_W\left(\frac{2}{\sqrt3}y_n
-\frac{1}{\sqrt6}z_n-\frac{2}{3}w_nt\right),
\label{gim3c}
\end{equation}
\end{mathletters}
\begin{equation}
R^n_{d_1}=R^n_{d_\alpha}=R^n_{d'_i}=\frac{2}{3}\,c_W\,w_n\,t,
\label{gim4}
\end{equation}
for the charge $-1/3$ quarks. We have checked numerically that only
for $n=1$ we have (for a given value of $t$ and $a<0.1$)
\begin{mathletters}
\label{ls}
\begin{equation}
L^1_{u_1}=L^1_{u_2}=L^1_{u_3}\not= L^1_{u'},
\label{lu}
\end{equation}
and
\begin{equation}
L^1_{d_1}=L^1_{d_2}=L^1_{d_3}\not=L^1_{d'_1}=L^1_{d'_2}.
\label{ld}
\end{equation}
\end{mathletters}
Hence, we can introduce a discrete symmetry, as in Model I of
Ref.~\cite{mpp}, in order to obtain a mass matrix which does not mix
$u_k$ with $u'$  and $d_k$ with $d'_i$, $k=1,2,3$ and $i=1,2$. That
is, the mass matrices have a tensor product form, next they can be
diagonalized with unitary matrices which are themselves tensor
products of unitary matrices. We see
that in this case the GIM mechanism~\cite{gim} is implemented, at
tree level, in the
$Z_1(\equiv Z^0)$ couplings. We must stress that, if the new quarks
$u'$ and $d'_i$ are very heavy, the requirements for natural (independent
of mixing angles) flavor conservation in the neutral currents to
order $\alpha G_F$~\cite{gw} break down, and it should be necessary to
impose the restriction that the mixing angles between ordinary and
the new heavy quarks must be very small~\cite{poggio}.

It is useful to define the coefficient $V=(L+R)/2$ and $A=(L-R)/2$.
In the Standard Model, at tree level, we have
$V_\psi^{SM}=t_{3L_\psi}-2Q_\psi\sin^2\theta_W$ and
$A^{SM}_\psi=t_{3L_\psi}$, where $t_{3L_\psi}$ is the weak isospin of the
fermion $\psi$. Hence, we have $V^{SM}_U\approx0.19$ and $A^{SM}_U=0.5$
for the charge 2/3 sector, and $V^{SM}_D\approx-0.345$ and
$A^{SM}_D=-0.5$ for the charge $-1/3$ sector. In our model,
also at tree level, using $a=0.01$ and $t=1.79$ we obtain
$V_U\approx0.19$, $A_U\approx0.5$ for $u_k$; and $V_D\approx-0.345$,
$A_D\approx-0.5$ for $d_k$. We see that the values are in
agreement with the values of the Standard Model. On the other hand
$V_{u'}\approx-0.310$, $A_{u'}\approx0$ and $V_{d'_i}\approx0.15$,
$A_{d'_i}\approx0$.

For leptons we have
\begin{equation}
L^n_\nu=L^n_{u_1}+\frac{4}{3}c_Ww_nt,R^n_\nu= -L^n_{u'}-\frac{4}{3}c_Ww_nt,
\quad
L^n_l=L_{d_1}+\frac{4}{3}c_Ww_nt,\quad R^n_l=-\frac{3}{\sqrt6}z_n.
\label{leptons}
\end{equation}

For $a=0.01$ and $t=1.79$ we obtain $V_\nu\approx0.5$,
$A_\nu\approx0.5$ and
$V_l\approx-0.036 $, $A_l\approx-0.5$ which are also in agreement with
the values of the Standard Model, $V^{SM}_\nu=A^{SM}_\nu=0.5$
and $V^{SM}_l\approx-0.035$
$A^{SM}_l=-0.5$.  Notice
also that at tree level neutrinos are still massless but they will get
a calculable mass through radiative corrections. In this kind of
model it is possible to implement
the Voloshin's mechanism i.e., in the limit of exact symmetry, a
magnetic moment for the neutrino is allowed, and a mass is
forbidden~\cite{voloshin}. We will not discuss this issue here.

Finally, we write down the charged current interactions in terms of
the symmetry eigenstates. In the
leptonic sector they are
\begin{equation}
{\cal L}^{CC}_l=-\frac{g}{2}\left[\bar\nu_L\gamma^\mu l_LW^+_\mu+
\overline{\nu^c_L}\gamma^\mu l_LV^+_{1\mu}+\overline{l^c_L}\gamma^\mu
\nu^c_LV^+_{2\mu}+\overline{l^c_L}\gamma^\mu
l_LU^{++}_\mu\right]+H.c.
\label{ccl}
\end{equation}
We have also the interaction $(g/2)\bar\nu^c_L\gamma^\mu\nu_LX^0$.

In the quark sector we have
\begin{eqnarray}
{\cal L}^{CC}_Q&=&-\frac{g}{\sqrt2}\left[\bar u_{k L}\gamma^\mu d_{k
L}W^+_\mu+\left(\bar u'_L\gamma^\mu d_{1L}+\bar u_{\alpha L}\gamma^\mu
d'_{iL} \right)V^+_{1\mu}\right.\nonumber \\ & &\mbox{}
+\left.\left( \bar J_L\gamma^\mu u_{1L}+\bar d_{\alpha L}\gamma^\mu j_{iL}
\right)V^+_{2\mu}+\left(\bar J_L\gamma^\mu u'_L\right.\right.
\nonumber \\ & &\mbox{}+\left.\left.\bar d'_{iL}\gamma^\mu j_{iL}\right)
V^+_{3\mu}+\left(\bar J_L\gamma^\mu d_{1L} - \bar u_{iL}
\gamma^\mu j_{iL}\right)U^{++}_\mu\right]+H.c.,
\label{ccq}
\end{eqnarray}
where $k=1,2,3$; $\alpha=2.3$; $i=1,2$. We have also interaction via
$X^0$ among primed and unprimed quarks of the same charge as $\bar
u'_L\gamma^\mu u_L$ and so on.

\acknowledgements

We would like to thank the
Con\-se\-lho Na\-cio\-nal de De\-sen\-vol\-vi\-men\-to Cien\-t\'\i
\-fi\-co e Tec\-no\-l\'o\-gi\-co (CNPq) full (FP) and partial (VP)
financial support. We  also would like to thank M.D. Tonasse for
useful discussions.


\begin{references}
\bibitem{fn} This value for $t$ is chosen in order to obtain the
right value for the neutral coupling for quarks in (\ref{gim1})-(\ref{gim4}).
\end{references}
\end{document}